\documentclass[aps,twocolumn,prd,preprintnumbers,superscriptaddress,nofootinbib, floatfix]{revtex4-1}  

\usepackage[linktocpage,pagebackref=false,hidelinks]{hyperref}

\usepackage{amsmath,setspace,amsfonts,latexsym}
\usepackage{amssymb}
\usepackage{color}
\usepackage{epsfig}
\usepackage{graphicx}
\usepackage{slashed}

\definecolor{Purple}{rgb}{0.637,0.285,0.641}

\def\bea{\begin{eqnarray}}
\def\eea{\end{eqnarray}}

\def\beq{\begin{equation}}
\def\eeq{\end{equation}}

\newcommand{\lsim}{\lesssim}
\newcommand{\gsim}{\gtrsim}

\def\ba{\begin{array}}
\def\ea{\end{array}}

\newcommand{\keV}{{\rm keV}}

\newcommand{\MeV}{{\rm MeV}}
\newcommand{\GeV}{{\rm GeV}}
\newcommand{\TeV}{{\rm TeV}}
\newcommand{\nc}{n_{\rm c}}
\newcommand{\mAd}{m_{A'}}

\begin{document}

\preprint{KEK--TH--2255}

\title{A solar origin of the XENON1T excess without stellar cooling problems}

\author{Sabyasachi Chakraborty}
\email{sabya@hep.fsu.edu}
\affiliation{Department of Physics, Florida State University, Tallahassee, FL 32306, USA}

\author{Tae Hyun Jung}
\email{thjung0720@gmail.com}
\affiliation{Department of Physics, Florida State University, Tallahassee, FL 32306, USA}

\author{Vazha Loladze}
\email{vloladze@fsu.edu}
\affiliation{Department of Physics, Florida State University, Tallahassee, FL 32306, USA}

\author{Takemichi Okui}
\email{tokui@fsu.edu}
\affiliation{Department of Physics, Florida State University, Tallahassee, FL 32306, USA}
\affiliation{High Energy Accelerator Research Organization (KEK), Tsukuba 305-0801, Japan}

\author{Kohsaku Tobioka}
\email{ktobioka@fsu.edu}
\affiliation{Department of Physics, Florida State University, Tallahassee, FL 32306, USA}
\affiliation{High Energy Accelerator Research Organization (KEK), Tsukuba 305-0801, Japan}

\begin{abstract}
Solar interpretations of the recent XENON1T excess events, such as axion or dark photon emissions from the sun, are thought to be at odds with stellar cooling bounds from the horizontal branch stars and red giants. We propose a simple effective field theory of a dark photon in which a $Z_2$ symmetry forbids a single dark photon emission in the dense stars, thereby evading the cooling bounds, while the $Z_2$ is spontaneously broken in the vacuum and sun, thereby explaining the XENON1T excess. The scalar responsible for the $Z_2$ breaking has an extremely flat potential, but the flatness can be maintained under quantum corrections. The UV completion of the EFT generally requires the existence of new electrically charged particles with sub-TeV masses with $O(1)$ couplings to the dark photon, offering the opportunity to test the scenario further and opening a new window into the dark sector in laboratory experiments.
\end{abstract}

\maketitle

\section{Introduction}
\label{sec:intro}
The XENON1T experiment has recently reported excess events with more than 3$\sigma$ significance~\cite{Aprile:2020tmw}. An attractive explanation is that a new light particle, such as an axion-like particle (ALP) or a dark photon, with a coupling to the electron is produced in the Sun and absorbed by the XENON detector~\cite{An:2014twa}.  Ref.~\cite{Aprile:2020tmw} also explores other scenarios: bosonic dark matter, non-standard neutrino interaction, and tritium concentration. 

The solar interpretations of the excess, however, are inconsistent with various stellar cooling bounds~\cite{Raffelt:1996wa, Redondo:2008aa, An:2013yfc, Giannotti:2015kwo, Hardy:2016kme, Giannotti:2017hny},
because the very interaction that produces the new particle at the sun and allows it to be detected by the XENON1T experiment also produces it in other stars such as horizontal branch (HB) stars and red giants (RG)\@.
This would increase the cooling rates of those stars and contradict with observations~\cite{An:2020bxd, Gao:2020wer,  Aprile:2020tmw,DiLuzio:2020jjp,Bloch:2020uzh,Budnik:2020nwz,Athron:2020maw}.

\begin{figure}[t]
 \begin{center}
  \hspace*{-1em}
  \includegraphics[width=0.47\textwidth]{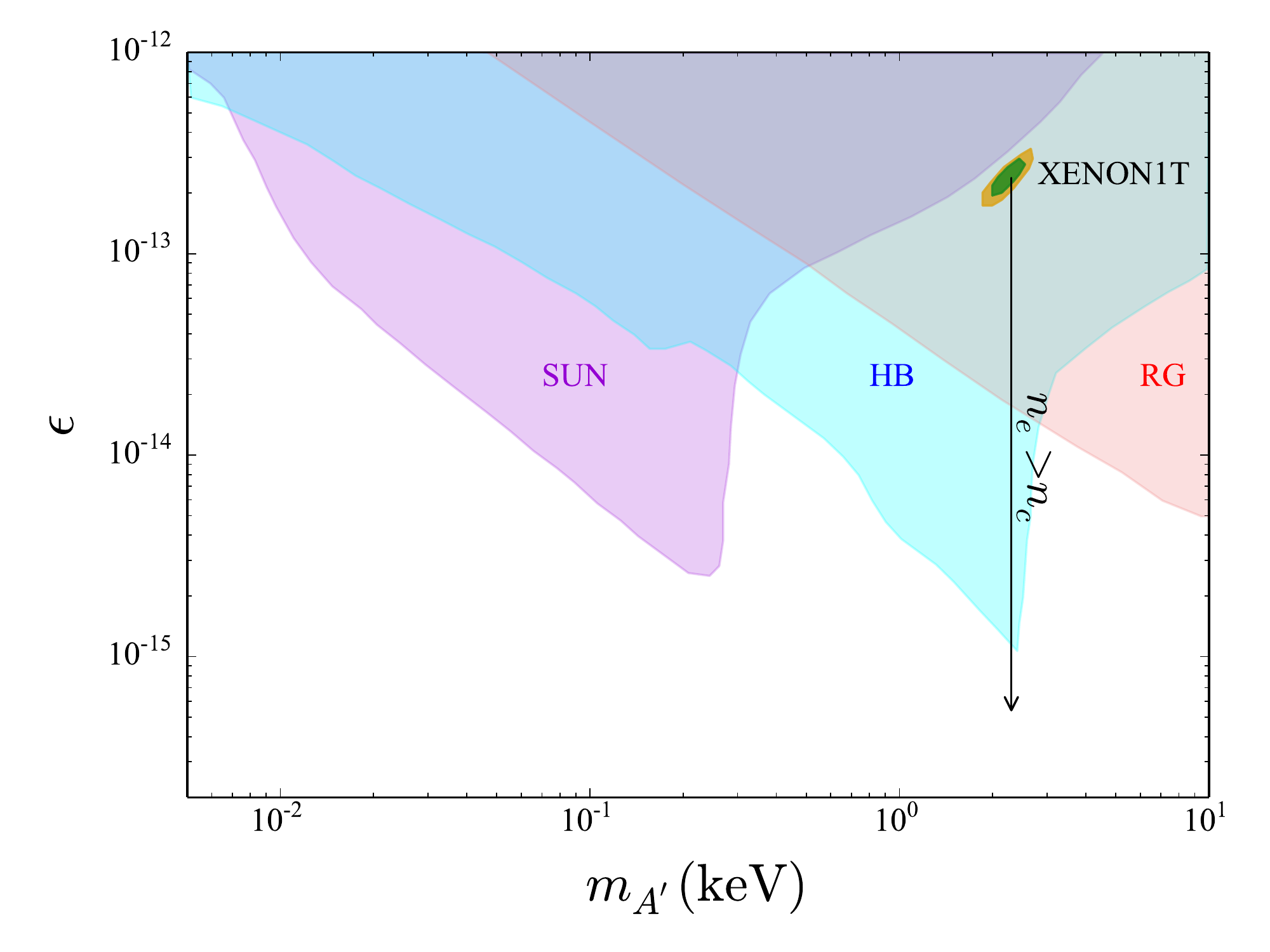}
\caption{An illustration of our scenario. In the vacuum/sun, the dark photon mass, $\mAd$, and its kinetic mixing with the photon, $\epsilon$, take the values matching the XENON1T excess, given by the small oval~\cite{Bloch:2020uzh}.
Inside HB/RG star cores, the electron density $n_e$ exceeds a critical value $\nc$ and $\epsilon$ vanishes, as represented by the arrow, evading the HB/RG bounds \cite{An:2014twa}. 
}
 \label{fig:bestfit}
 \end{center}
 \end{figure}

In this paper, we propose a solar interpretation of the XENON1T excess with a dark photon that can avoid the stellar cooling constraints.
Our scenario is illustrated in Fig.~\ref{fig:bestfit},
where the two axes corresponds to the dark photon mass, $\mAd$, and the kinetic mixing, $\epsilon$, between the dark and SM photons.
In the vacuum as well as in the sun, 
$\mAd$ and $\epsilon$ take the values that match the XENON1T excess~\cite{Bloch:2020uzh}, represented by the small oval.
On the other hand, in a denser environment such as the core of a HB or RG star, where the electron number density $n_e$ exceeds a critical value $\nc$, 
$\epsilon$ vanishes and takes us out of the stellar cooling bounds, as indicated by the downward arrow.

We realize this scenario by promoting $\epsilon$ to a dynamical operator $\phi(x) / \Lambda'$ with some mass scale $\Lambda'$ and a scalar field $\phi$ whose expectation value depends $n_e$.
This is achieved by having the mass-squared parameter of $\phi$ depend on $n_e$ as $m^2 = -\mu_\phi^2 + m_e \bar{e} e / \Lambda^2$, where $\mu_\phi$ and $\Lambda$ are chosen such that $\nc$ falls between the solar and HB densities.%
\footnote{The operator $\bar{e} e$ is not exactly equal to the electron number density, $\bar{e} \gamma^0 e$,
but they are nearly equal for nonrelativistic electrons with velocities $v \ll 1$ as the difference between them are $O(v^2)$. We ignore this difference throughout our analysis.}

This idea will be fully developed in Sec.~\ref{sec:dark} as an effective field theory (EFT) and its phenomenological constraints as well as theoretical consistency will be studied.
In particular, while obtaining the right phenomenology requires the tree-level potential for $\phi$ to be extremely flat, we will show that the flatness is nonetheless maintained even under loop corrections, 
except for the fine tuning of the $\phi$ mass as expected for the mass of an elementary scalar field not protected by a symmetry.
We believe this is the first model in the literature on a density-dependent solar interpretation of the XENON1T anomaly without the cooling problems that does not suffer from too large quantum corrections destroying the flatness of the potential.
As for other directions for a solar interpretation of the excess,  it is also possible to exploit {\it composition difference} (i.e., H versus He) between the sun and denser stars instead of density differences~\cite{McKeen:2020vpf}.

We will discuss the UV completion of the EFT in Sec.~\ref{sec:UV}, where we present a general argument that the allowed range of parameters in the EFT necessarily point to the existence of new electrically charged particles with sub-TeV masses in any UV-complete realization of the EFT\@.
This offers unexpected additional experimental probes into our scenario by laboratory experiments.   

While our scenario is not the only way to use density dependence,
we will discuss advantages of our scenario over other similar ideas as well as earlier work on density dependent interactions in the context of solar interpretation of the XENON1T excess~\cite{Bloch:2020uzh, DeRocco:2020xdt} in Sec.~\ref{sec:difficulties}, focusing on the stability of the flat potential under quantum corrections.

\section{The Dark Photon Model}
\label{sec:dark}
In this section, we provide a simple explicit model that realizes the scenario described in Sec.~\ref{sec:intro}.
We begin by an effective field theory (EFT) that is self-consistent and suffices to study phenomenological implications of our scenario.
The UV completion of the EFT will be discussed in Sec.~\ref{sec:UV}.

\subsection{The effective Lagrangian}
The EFT introduces two new degrees of freedom beyond the SM: a massive spin-1 field, $A'_\mu$, and a real scalar field, $\phi$.
The tree-level effective Lagrangian is given by
\begin{align}
{\cal L} 
= {\cal L}_\text{SM} 
+ {\cal L}_\text{kin}^{A'\phi}
+ \frac{\mAd^2}{2} A'^2
- \frac{\phi}{2\Lambda'} B^{\mu\nu} F'_{\mu\nu}
- V(\phi)
\,,\label{eq:Lagrangian}
\end{align}
where ${\cal L}_\text{kin}^{A'\phi}$ is just the canonically normalized kinetic terms of $A'$ and $\phi$, while $B_{\mu\nu}$ and $F'_{\mu\nu}$ are the field strength tensors for the SM hypercharge gauge field and $A'$, respectively.
It is essential here that ${\cal L}_\text{kin}^{A'\phi}$ does {\it not} contain the ``bare'' kinetic mixing term, $B^{\mu\nu} F'_{\mu\nu}$, as we want the mixing to be entirely induced by $\langle \phi \rangle$.
The absence of the bare mixing is consistent within the EFT 
as the Lagrangian~\eqref{eq:Lagrangian} has the following $Z_2$ symmetry:
\bea
\phi \to -\phi
\quad\text{and}\quad
A'_\mu \to -A'_\mu
\label{eq:Z_2}
\eea
with all other fields unchanged.
A possible UV origin of this $Z_2$ symmetry will be discussed in Sec.~\ref{sec:UV}. 
The dependence of $\langle \phi \rangle$ on the electron density comes from
\bea
V(\phi) 
= \frac{1}{2} \!\left( 
- \mu_\phi^2 
+ \frac{y_e H \bar{\ell}_{\rm L} e_{\rm R} + \text{h.c.}}{\Lambda^2} 
\right)\! \phi^2 + \frac{\lambda_\phi}{4!} \phi^4
\,,\label{eq:Vphi}
\eea
where, upon electroweak symmetry breaking, the numerator of $\Lambda^2$ picks up an electron density dependence:
\begin{align}
y_e \langle H \rangle \bar{\ell}_{\rm L} e_{\rm R} + \text{h.c.}  
= m_e \bar{e} e
\,.
\end{align}

In the vacuum and the core of the sun,
we would like the the density-dependent term, $m_e \bar{e} e / \Lambda^2$, to be negligible compared to $\mu_\phi^2$ in the potential~\eqref{eq:Vphi}.
In these environments, we have $\langle \phi \rangle = v_0 \neq 0$ and
an effective $\gamma$-$A'$ mixing is induced: 
\bea
{\cal L}_\text{induced} = -\frac{\epsilon}{2} F^{\mu\nu} F'_{\mu\nu}
\quad\text{with}\quad
\epsilon \equiv \frac{v_0 \cos\theta_w}{\Lambda'}
\,,
\label{eq:mixing}
\eea
where $\theta_w$ is the electroweak mixing angle.
The best fit values of $\mAd$ and $\epsilon$ to the XENON1T excess were found by Ref.~\cite{Bloch:2020uzh} to be
\bea
\mAd = 2.3\,\keV
\,,\quad
\epsilon = 2.5 \times 10^{-13}\,.
\label{eq:XENON1T_best_fit}
\eea

On the other hand, in denser stars, we would like the density-dependent term to dominate over $\mu_\phi^2$ 
so that we have $\langle \phi \rangle = 0$ and thereby turn off $\gamma$-$A'$ mixing. 
Let $\nc$ be a critical density at which the coefficient of $\phi^2$ in the potential~\eqref{eq:Vphi} changes the sign.
Since the electron number densities in the cores of the sun and a typical HB star are respectively $\sim 6 \times 10^2\>\keV^3$ and $\sim 4 \times 10^4\>\keV^3$~\cite{Hardy:2016kme}, 
$\nc$ should be between these densities, e.g., $4 \times 10^3\>\keV^3$.
We then have
\bea
\mu_\phi
= \frac{\sqrt{m_e \nc}}{\Lambda}
\,.\label{eq:mu_phi}
\eea
Then, from $\lambda_\phi=6\mu_\phi^2/v_0^2$
 with the relation~\eqref{eq:mixing} to eliminate $v_0$,
this is equivalent to
\begin{align}
\lambda_\phi = \frac{6 m_e \nc \cos^{2\!}\theta_w}{\epsilon^2 \Lambda^2 \Lambda'^2}
\,.\label{eq:lambda_phi}
\end{align}
From Eqs.~\eqref{eq:mu_phi} and~\eqref{eq:lambda_phi}, we will see the generic feature of our parameter space that the values of $\mu_\phi$ and $\lambda_\phi$ tend to be very small because it will turn out that $\Lambda'\gtrsim10^9\,\GeV$ and $\Lambda\gtrsim10^7\,\GeV$.

\subsection{Phenomenological constrants}
We now discuss constraints on our model.
While the HB/RG bounds on the dark photon are evaded by construction, 
we must require $\phi$ not to be produced in stars.
This requirement must be analyzed separately for the sun and the HB/RG cases
as $\langle \phi \rangle = v_0$ for the former while $\langle \phi \rangle = 0$ for the latter. 

\begin{figure}[t!]
 \begin{center}
  \includegraphics[width=0.5\textwidth]{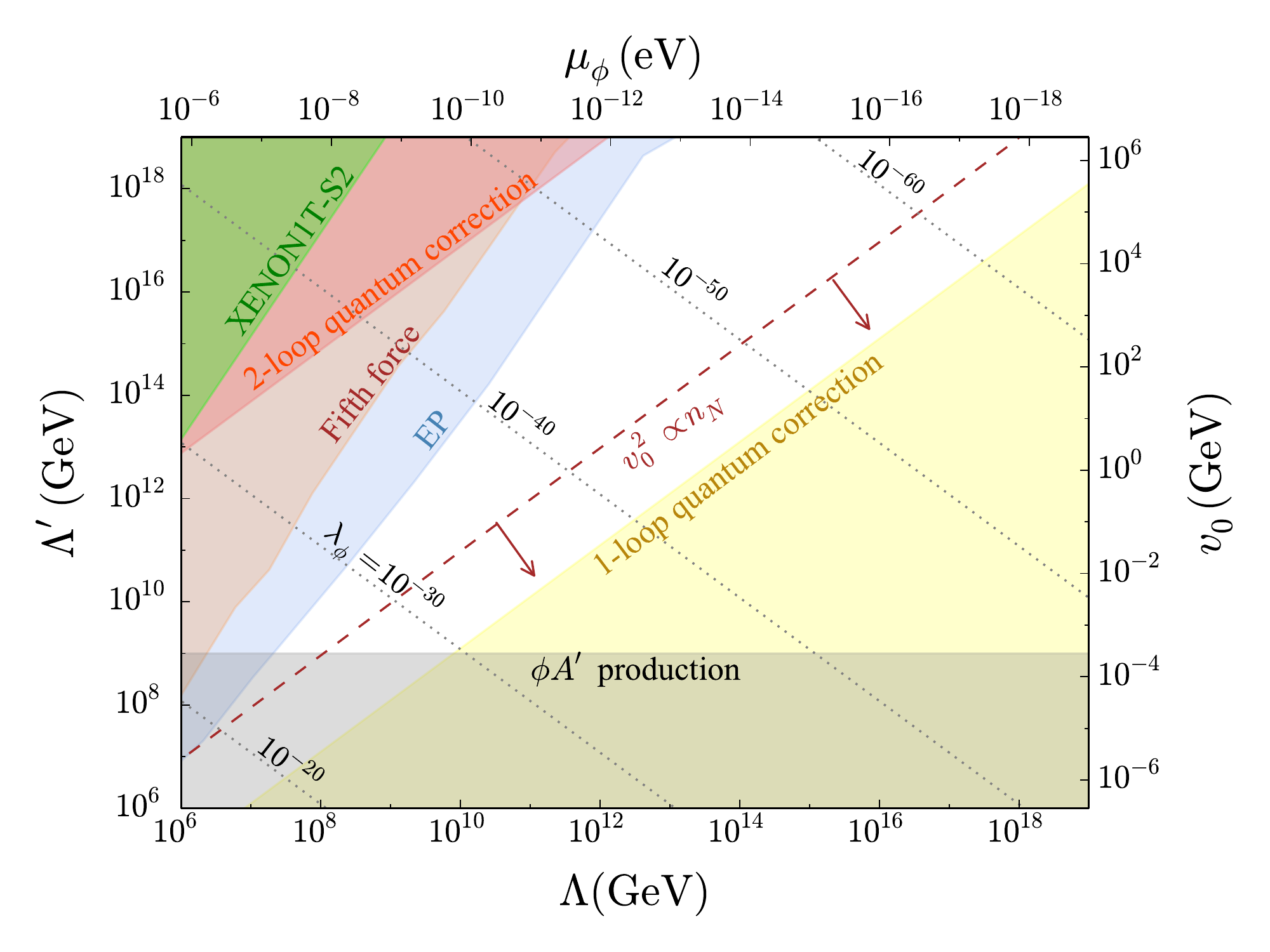}
 \end{center}
 \vspace{-10pt}
\caption{Constraints on the parameter space of the dark photon model of Sec.~\ref{sec:dark} for $\epsilon = 2.5 \times 10^{-13}$, $\nc = 4 \times 10^3~\keV^3$, and $Y=2$.
The shaded region in green is excluded by the XENON1T S2-only analysis (Eq.~\eqref{phi_emission_constraint}), 
while the regions in brown and blue by fifth force and equivalence principle (EP) tests, respectively. 
The red and yellow regions suffer from too large quantum corrections (Eqs.~\eqref{eq:twoloopconst} and \eqref{eq:gaugebosonloop}, respectively).
The gray region shows a conservative bound from $\phi A'$ production (Eq.~\eqref{eq:phiAprod}).
Below the dashed brown line, $\langle\phi\rangle$ would be controlled by the nucleon density instead of the electron density (Eq.~\eqref{eq:nucleon}).
The dotted lines show the values of $\lambda_\phi$ (Eq.~\eqref{eq:lambda_phi}).
}
\label{fig:moneyplot}
\end{figure}

In the sun, we have $\phi = v_0 + \varphi$, where $\varphi$ can be singly produced through an effective Yukawa interaction with electrons, $y_{\rm eff} \, \varphi \bar{e} e$, with $y_{\rm eff} = m_e v_0 / \Lambda^2$.
Such emission from the sun is constrained by XENON1T S2-only analyses~\cite{Budnik:2019olh}, and it 
gives a strong
bound on $y_{\rm eff}$ in the $\langle \phi \rangle = v_0$ case.
There is a stronger bound by RG star cooling~\cite{Hardy:2016kme} 
but it is irrelevant here as it is derived from denser stars in which $\langle \phi \rangle = 0$.
Parametrizing the solar bound as $y_{\rm eff} < 10^{-15} \,Y$ (with $Y \simeq 2$ currently from Ref.~\cite{Budnik:2019olh}),
we obtain
\begin{align}
\frac{\Lambda'}{\Lambda^2} < 
7\,Y \,
\frac{2.5\times10^{-13}}{\epsilon}
\>\GeV^{-1}
\,.\label{phi_emission_constraint}
\end{align}
We use this relation between $\Lambda$ and $\Lambda^\prime$ to rule out the area shown by the green region in Fig.~\ref{fig:moneyplot}.

On the other hand, in a dense star like HB or RG, $\langle \phi \rangle$ vanishes so the unbroken $Z_2$ requires $\phi$ to be produced
always in pairs or in association with $A'$.
For the $\phi$ pair production,
compared to the single $\varphi$ production case above, the presence of an extra $\phi$ in the final state implies that 
the $\phi$ pair production rate is roughly given by the single $\varphi$ production rate with $y_{\rm eff}$ replaced with $y'_\text{eff} \equiv m_e T/(4\pi\Lambda^2)$, 
where $T$ is the core temperature of the star ($T \sim 10\>\keV$ for a typical HB/RG)\@. 
With the strongest bound $y'_\text{eff} <7\times10^{-16}$ from RG stars~\cite{Hardy:2016kme},
we get $\Lambda\gtrsim$~TeV but 
we will see that there are much stronger constraints from fifth force/equivalence principle tests.

Similarly, the emission of $\phi$-$A'$ pairs can contribute to stellar cooling. 
We can estimate the $\phi$-$A'$ production rate $\Gamma_{\phi A'}$ as
\begin{equation}
    \Gamma_{\phi A'}(m_{A'})\sim \frac{1}{16\pi^2}\frac{T^2}{\Lambda'^2} \frac{\Gamma_{A'}(M_{\phi A'})}{\tilde{\epsilon}^2} \,,
\end{equation}
where $T$ is the temperature of the star, while $\Gamma_{A'}(M_{\phi A'})$ is the would-be production rate of a single dark photon with mass $M_{\phi A'}$ and mixing $\tilde{\epsilon}$, where $M_{\phi A'}$ is the invariant mass of the $\phi$-$A'$ system. 
To be conservative, we take  $M_{\phi A'}\sim m_{\rm peak}$ with $m_{\rm peak}$ being where the cooling rate is maximized in Fig.~1 of  Ref.~\cite{An:2020bxd}. We then get
\begin{equation}
\frac{T_{\rm HB/RG}}{4\pi\Lambda'}\lesssim 6\times10^{-16} 
~~\Rightarrow~~
\Lambda'\gtrsim10^9\,\GeV
\,.
\label{eq:phiAprod}
\end{equation}
This excludes the horizontal grey region in Fig.\,\ref{fig:moneyplot}

Because of the large lower bound of $\Lambda$, $\phi$ needs to be very light (see Eq.\,\eqref{eq:mu_phi}).
Such light $\phi$ provides an additional long-range force, and we find that fifth force experiments and equivalence principle tests give the strongest bound on $y_{\rm eff}$ and $m_\phi=\sqrt{2}\mu_\phi$ of our model.
We convert the bounds obtained in Ref.\,\cite{Wise:2018rnb,Kapner:2006si,Salumbides:2013dua, Schlamminger:2007ht} to our parameters (see the brown and blue regions in Fig.\,\ref{fig:moneyplot}).

In addition, the mass of a light scalar field with a small self interaction can be constrained by black hole superradiance.
However, based on the analysis of Ref.\,\cite{Arvanitaki:2014wva}, our self-coupling $\lambda_\phi$ is not small enough to maintain  superradiance, which requires $\lambda_\phi \lsim 10^{-65}$.
Therefore, black hole superradiance does not give a relevant constraint on our scenario.

Next, the presence of the $\bar{e}e\phi^2$ term in the potential~\eqref{eq:Vphi} predicts that the electron mass depends on $\langle \phi \rangle$. 
In an environment denser than $\nc$, 
the electron mass 
changes as
\begin{align}
\frac{m_e^{\rm dense} - m_e}{m_e}
&= -\frac12 \Bigl( \frac{v_0}{\Lambda} \Bigr)^{\!2}
\nonumber\\
&\simeq
-4 \times 10^{-26}
\!\left( \frac{\Lambda'}{\Lambda} \right)^{\!2} 
\!\left( \frac{\epsilon}{2.5\times 10^{-13}} \right)^{\!2}
.\label{eq:electronmasscorrection}
\end{align}
In the sun, the deviation is further suppressed by a factor of ${n_e^\odot}/{\nc}$, where $n_e^\odot \sim 10^2\>\keV^3$.
This is an unobservably small deviation because the fifth-force/equivalence principle bounds prevent us from enhancing the deviation to an observable level by having a large hierarchy between $\Lambda'$ and $\Lambda$.

\subsection{Constraints from radiative corrections}
Finally, as we can see in the small values of $\mu_\phi$ and $\lambda_\phi$ in Fig.~\ref{fig:moneyplot}, we need a very flat potential for $\phi$,
which requires us to check that the flatness is not destroyed by loop corrections.
While the coefficient of $\phi^2$ needs to be tuned as expected for any elementary scalar whose mass is not protected by any symmetry,
we would like to avoid having to keep tuning
the coefficients of $\phi^4$, $\phi^6$, $\phi^8$, \ldots, to maintain the flatness of the potential at loop level.

First, we  investigate the 1-loop Coleman-Weinberg (CW) potential induced from $\gamma/Z/A'$ loops.
The largest contributions come from integrating out $Z$ and are given in dimensional regularization by
\begin{equation}
\Delta V_g 
= \frac{3}{64\pi^2}(M_Z^2)^2 \, \ln{\frac{M_Z^2}{\mu^2}}
\,,\label{eq:CWgaugeboson}
\end{equation}
where  we have absorbed the scheme-dependent constant term in the normalization of $\mu$. 
The field dependent $Z$ mass, in the limit of neglecting $\mAd$ with respect to $m_Z$ but to all orders in $\phi / \Lambda'$,  is given by 
\begin{equation}
M_Z^2 
= m_Z^2 \!\left( 
1 + 
\frac{(\phi / \Lambda')^2}
     {1 - (\phi / \Lambda')^2} 
\sin^2\!\theta_w
\right),
\label{eq:CWgaugebosonmass}
\end{equation}
where $m_Z$ is the SM $Z$ mass. 
It is clear from the form of this CW potential that it suffices to require that the correction to the quartic coupling $\lambda_\phi$ should be at most $\lambda_\phi$ itself, as the successive higher order terms are suppressed by powers of $v_0^2 / \Lambda'^2\sim \epsilon^2\ll1$. 
Thus, taking into account scale uncertainties, we roughly impose
\begin{equation}
\frac{\lambda_{\phi}}{4!}
\gtrsim
\frac{\sin^2\!\theta_w}{16\pi^2}
\!\left( \frac{m_Z}{\Lambda'} \right)^{\!4}.
\label{eq:CWgauge}
\end{equation}
Using the relation~\eqref{eq:lambda_phi} to eliminate $\lambda_\phi$, this becomes
\begin{align}
\frac{\Lambda'}{\Lambda} 
\gsim  0.1 \>
\frac{\epsilon}{2.5\times10^{-13} }
\sqrt{\frac{4 \times 10^3\>\keV^3}{\nc}}
\,.\label{eq:gaugebosonloop}    
\end{align}
The constraint derived from Eq.~(\ref{eq:gaugebosonloop}) excludes the yellow region in Fig.~\ref{fig:moneyplot}.

\begin{figure}[t!]
 \begin{center}
  \includegraphics[width=0.48\textwidth]{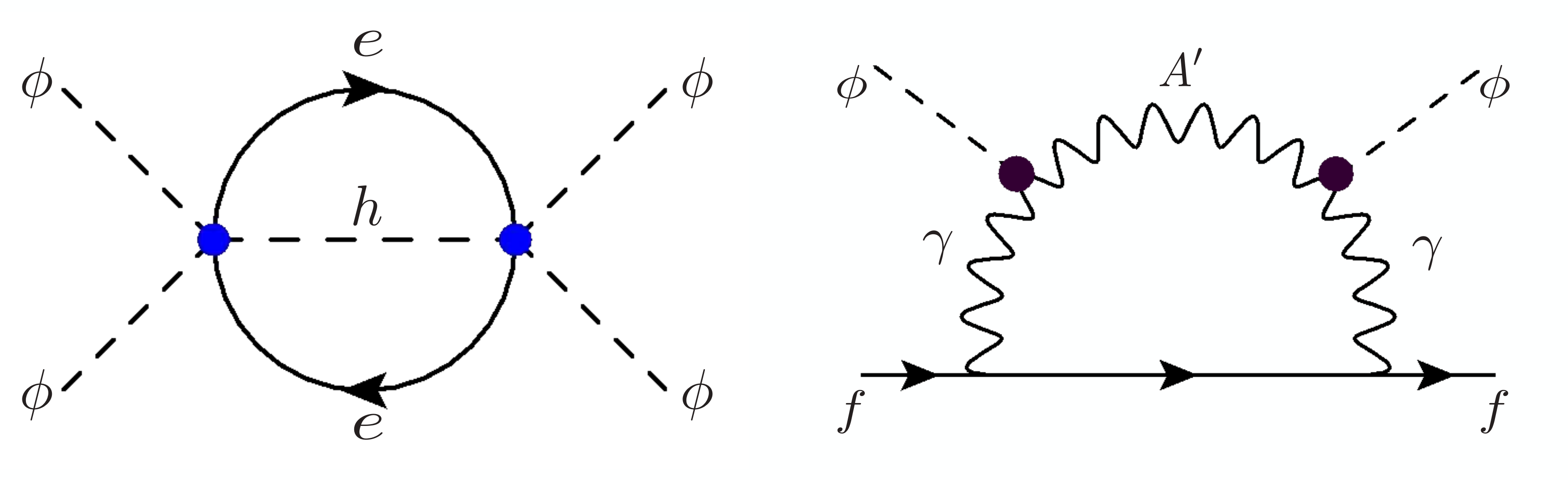}
 \end{center}
\caption{
Examples of diagrams contributing to Eq.~\eqref{eq:e-h-loop}~(left) and Eq.~\eqref{eq:fermion}~(right). 
}
\label{fig:e-h-diagram}
\end{figure}

Secondly, we  consider corrections due to the $m_e \bar{e} e / \Lambda^2$ term in the tree-level potential~\eqref{eq:Vphi}. 
The 1-loop CW potential for $\phi$ from an electron loop is given by
\begin{eqnarray}
\Delta V_e 
= -\frac{M_e^4}{8\pi^2} \, \ln{\frac{M_e}{\mu}}
\,,\quad
M_e = m_e \!\left( 1 + \frac{\phi^2}{2\Lambda^2} \right). 
\label{eq:CWelectron}
\end{eqnarray}
We see again that it suffices to ensure that corrections to $\phi^4$ are less than $\lambda_\phi$ in the relevant range of $\phi$, because $v_0 / \Lambda  \ll 1$.
This roughly amounts to 
\begin{equation}
\frac{\lambda_\phi}{4!}
\gtrsim
\frac{1}{8\pi^2} \!\left(\frac{m_e}{\Lambda} \right)^{\!4}
.\label{eq:CWeconstraint}
\end{equation}
Here, $1 / \Lambda$ is only multiplied by a small mass $m_e$, while $1 / \Lambda'$ is multiplied by a large mass $m_Z$  in the other 1-loop constraint~\eqref{eq:CWgauge}.  
Therefore, we could get a stronger constraint on $\Lambda$ from a 2-loop diagram where $1/\Lambda$ comes with a large mass such as $m_{Z/W}$ or $m_h$. 
A diagram depicted in Fig.~\ref{fig:e-h-diagram} left, for example, is enhanced by $m_h$.
We estimate the bound from this as
\bea
\lambda_\phi
\gtrsim
\frac{y_e^2}{16\pi^4}
\!\left(\frac{m_h}{\Lambda}\right)^{\!4}
\sim
\frac{1}{16\pi^4} \frac{m_e^2 m_h^2}{\Lambda^4}
\,,\label{eq:e-h-loop}
\eea
where, to be conservative, we have only kept one power of 16 to account for a possible $O(10)$ numerical factor in the numerator. 
This is indeed much stronger than the 1-loop bound~(\ref{eq:CWeconstraint}).
(We also get similar bounds from $Z/W$ exchange instead of $h$.)
Using the relation~\eqref{eq:lambda_phi} to eliminate $\lambda_\phi$, the bound~\eqref{eq:e-h-loop} becomes
\begin{align}
\frac{\Lambda'}{\Lambda} 
\lsim 8 \times 10^6 \>
\frac{2.5\times10^{-13}}{\epsilon}
\sqrt{\frac{\nc}{4 \times 10^3\>\keV^3}}
\,.\label{eq:twoloopconst}    
\end{align}
Eq.~(\ref{eq:twoloopconst}) corresponds to the red region in Fig.~\ref{fig:moneyplot}.

In addition to the flatness of the potential, our model also has the property that it is the electron density, not the nucleon density, that controls $\langle \phi \rangle$.
This property is not necessary, and we could repeat the whole analysis for the nucleon case.
For simplicity, however, we require that the electron density plays the dominant role.
Starting from our tree-level Lagrangian~\eqref{eq:Lagrangian}, 
the operator $\bar{f} f \phi^2$ for fermion $f$ is induced at 1-loop from diagrams like Fig.~\ref{fig:e-h-diagram} right (each $\gamma$ can also be a $Z$).
This is given by
\begin{equation}
\sim \sum_f Q_f^2 \frac{\alpha_{\rm EM}}{\pi} \frac{m_f \bar{f} f}{\Lambda'^2} \frac{\phi^2}{2}
\,,\label{eq:fermion}
\end{equation}
where $m_f$ and $Q_f$ are the mass and electric charge of fermion $f$.
When $f$ is a light quark $q = u$, $d$, or $s$, the contribution from the QCD condensate, $m_q \langle \bar{q} q \rangle \sim m_\pi^2 f_\pi^2$, is absorbed in $\mu_\phi^2$, 
and the contributions from nucleons can be estimated by replacing $m_q \bar{q} q$ by $m_N \bar{N} N$ for nucleon $N$. 
Since $(\alpha_{\rm EM} / \pi) m_N / m_e \sim 4$, 
the requirement that the $m_e \bar{e} e / \Lambda^2$ operator should control $\langle \phi \rangle$ amounts to $\Lambda' \gsim$ a few$\times\Lambda$,
which is already stronger than the condition~\eqref{eq:gaugebosonloop},
but to be conservative we impose (shown by the dashed brown line in Fig.~\ref{fig:moneyplot})
\begin{equation}
\Lambda'\gtrsim 10\Lambda \,.
\label{eq:nucleon}
\end{equation}
Although this is not strictly necessary as we stated above, 
this and the bound~\eqref{eq:twoloopconst} provides the strongest conditions on the self-consistency of our analysis: 
\begin{equation}
10 \lsim \frac{\Lambda'}{\Lambda} \lsim 8\times10^{6}\;.
\end{equation}
Other potential constraints we checked such as top loop contributions and a radiatively generated $\phi^2 H^\dag H$ interaction do not impose additional restrictions on the parameter space.
We thus see a large parameter space in which quantum corrections do not destroy the flat potential (as well as the dominant role of the electron density).

\section{UV completion}
\label{sec:UV}

\subsection{General estimates of scales}
We begin by a general discussion that new degrees of freedom that UV-completes the EFT should appear roughly at the TeV scale.

Let $M'$ be a representative mass scale of new particles in the UV theory responsible for generating the $\phi B^{\mu\nu} F'_{\mu\nu}$ operator of the EFT\@.
To generate this operator,
the UV theory must contain an interaction linear in $\phi$.
Let $y'$ be the dimensionless coupling constant of that interaction (measured in units of $M'$ if it is dimensionful).
Then, roughly, we have
\begin{align}
    \frac{1}{\Lambda'} \sim \frac{g' g_Y}{4\pi^2} \frac{y'}{M'}
    \,,\label{eq:rough:matching:M'}
\end{align}
where $g'$ and $g_Y$ are the U(1)$_{A'}$ and U(1)$_Y$ gauge couplings, respectively,
and we have imagined a typical 1-loop factor, $(\text{several})/16\pi^2$.
(We will later see why it must be at 1-loop.)
On the other hand, using the same $y'$ vertex 4 times, we also generate a correction to the $\phi^4$ operator.
In order to avoid fine tuning in $\lambda_\phi$ upon matching the UV theory to the EFT, we must roughly impose
\begin{align}
    \frac{y'^4}{4\pi^2} \lsim \lambda_\phi
    \,.\label{eq:rough:no-tuning}
\end{align}
Since we know that $\lambda_\phi$ is extremely tiny, this shows that $y'$ must also be very small.
Then, in the relation~\eqref{eq:rough:matching:M'}, 
we see that $M$ can be at quite a low scale despite the fact that $\Lambda'$ is very high.
To proceed more concretely, we eliminate $\lambda_\phi$ in the condition~\eqref{eq:rough:no-tuning} by using the relation~\eqref{eq:lambda_phi} with $\epsilon = 2.5 \times 10^{-13}$ and $\nc = 4 \times 10^3\>\keV$.
This gives
\begin{align}
    y' \lsim \frac{O(10^2)\>\GeV}{\sqrt{\Lambda\Lambda'}}
    \,.\label{eq:rough:y}
\end{align}
Applying this inequality to the matching relation~\eqref{eq:rough:matching:M'}, we get
\begin{align}
    M' \lsim O(10^2) \, g' \sqrt{\frac{\Lambda' / \Lambda}{10^4}} \>\GeV
    \,,\label{eq:rough:M':Lambda-Lambda'}
\end{align}
where, from Fig.~\ref{fig:moneyplot}, the maximum possible value of $\Lambda' / \Lambda$ is roughly $O(10^4)$ in a large portion of the allowed region.

Let us do similar general estimates for the UV completion of the electron density dependent operator in~\eqref{eq:Vphi}.
Let $M$ be a representative mass scale of new particles responsible for generating the density dependent operator,
and $y$ be a representative size of the couplings of $\phi$ to those particles, defined such that each $\phi$ comes with one $y$
(that is, if the interaction is $\propto \phi^n$, its coefficient is  $\propto y^n$ by definition.)
Then, upon integrating out the new particles at $M$, we have
\begin{align}
    \frac{y_e}{\Lambda^2} = \frac{c y^2}{M^2}
\,,\label{eq:rough:matching:M}
\end{align}
where $c$ represents all other coupling constants and numerical factors in the diagram and, if not at tree level, loop factors.
As we estimated above for $y'$, 
avoiding fine tuning in $\lambda_\phi$ roughly requires 
$y \lsim O(10^2) \>\GeV / \sqrt{\Lambda\Lambda'}$.
Using this with the matching condition~\eqref{eq:rough:matching:M},
we get
\begin{align}
    M \lsim O(10^3)\>\GeV \> \sqrt{c} \, \sqrt{\frac{10^4}{\Lambda'/ \Lambda}} 
\,.\label{eq:rough:M:Lambda-Lambda'}
\end{align}
Both~\eqref{eq:rough:M':Lambda-Lambda'} and~\eqref{eq:rough:M:Lambda-Lambda'} point to a sub TeV scale for $M'$ and $M$, provided that $g' \sim c \sim O(1)$.

The existence of new particles with roughly sub-TeV masses is a {\it prediction} implied by the allowed region of the EFT parameters. 
Eq.~\eqref{eq:rough:M':Lambda-Lambda'} tells us that we must have $g ' \sim O(1)$ in order for the new particles at $M'$ not to have already been experimentally excluded, since they carry hypercharge to generate the $\phi B^{\mu\nu} F'_{\mu\nu}$ operator.
Therefore, it is guaranteed that we will have new electrically charged particles with sub-TeV masses with $O(1)$ couplings to the dark photon.
Moreover, the $\phi B^{\mu\nu} F'_{\mu\nu}$ operator should not be generated at a higher loop order than one as that would lower $M'$ too much.
This means the new particles at $M'$ must couple to $\phi$ at tree level,
albeit with tiny couplings.
Similarly, in order for the new particles at $M$ not to be too light, they must generate the density dependent operator at tree level with all couplings except $y$ being $O(1)$ so that $c \sim O(1)$.

Therefore, from the general estimates, we already see the opportunity to experimentally probe our scenario further with a window into the dark sector for laboratory experiments.
We also see that the structure of any explicit model of the UV completion is very constrained.

\subsection{An explicit UV model}
Now that it is testable experimentally, it is worth building an explicit example of the UV completion. 
Conceptually, we must also address the UV origin of the $Z_2$ symmetry~\eqref{eq:Z_2}, which we call $Z_2^{A'}$,
so let us begin with this.
Since $A'$ is odd under $Z_2^{A'}$ while all SM gauge fields are even,  
$Z_2^{A'}$ must act as charge conjugation for U(1)$_{A'}$
while it must not for the SM gauge group.
The simplest possibility, therefore, is to introduce two new heavy particles $\psi_{1,2}$ exchanged by $Z_2^{A'}$ with opposite U(1)$_{A'}$ charges and the same SM charges. The SM charges must at least include a nonzero hypercharge to generate the $\phi B^{\mu\nu} F'_{\mu\nu}$ operator. 
For simplicity and definiteness, we take $\psi_{1,2}$ to be Dirac fermions with $M'$, $Y_\psi$, and $\pm g'$ being their common mass, hypercharge, and the opposite U(1)$_{A'}$ charges, respectively.
For $\phi$, we have $\phi \to -\phi$ under $Z_2^{A'}$ in the UV just as in the EFT\@. 

Now the UV degrees of freedom and symmetries have been fixed.%
\footnote{If cosmological domain walls from the $Z_2^{A'}$ breaking become a problem, we can softly break $Z_2^{A'}$ by introducing a small mass splitting between $\psi_1$ and $\psi_2$.
This will induce a bare kinetic mixing between $B_{\mu\nu}$ and $F'_{\mu\nu}$, which must be smaller than the $10^{-15}$ level (see Fig.~\ref{fig:bestfit}).}
Gauge symmetries and $Z_2^{A'}$ allow just one coupling between $\psi_{1,2}$ and $\phi$:
\begin{align}
{\cal L} \supset -y' \, \phi (\bar{\psi_1} \psi_1 - \bar{\psi_2} \psi_2)
\,.\label{eq:y'_Yukawa}
\end{align}
Thus, the new particles $\psi_{1,2}$ couple to $\phi$ at tree level as required by the general analysis above.
This is another reason why we need at least two new degrees of freedom to realize $Z_2^{A'}$.
If we just had one $\psi$ and conjugate $\psi$ itself under $Z_2^{A'}$, we would not be able to couple it to $\phi$.

Accidentally, this UV model can also offer a WIMP DM candidate if $\psi_{1,2}$ contain an electrically neutral state, because the Lagrangian has an accidental U(1)$_{\psi_1}\times$U(1)$_{\psi_2}$ global symmetry, which ensures the neutral state to be stable if it is the lightest.

\begin{figure}[t!]
 \begin{center}
  \includegraphics[width=0.3\textwidth]{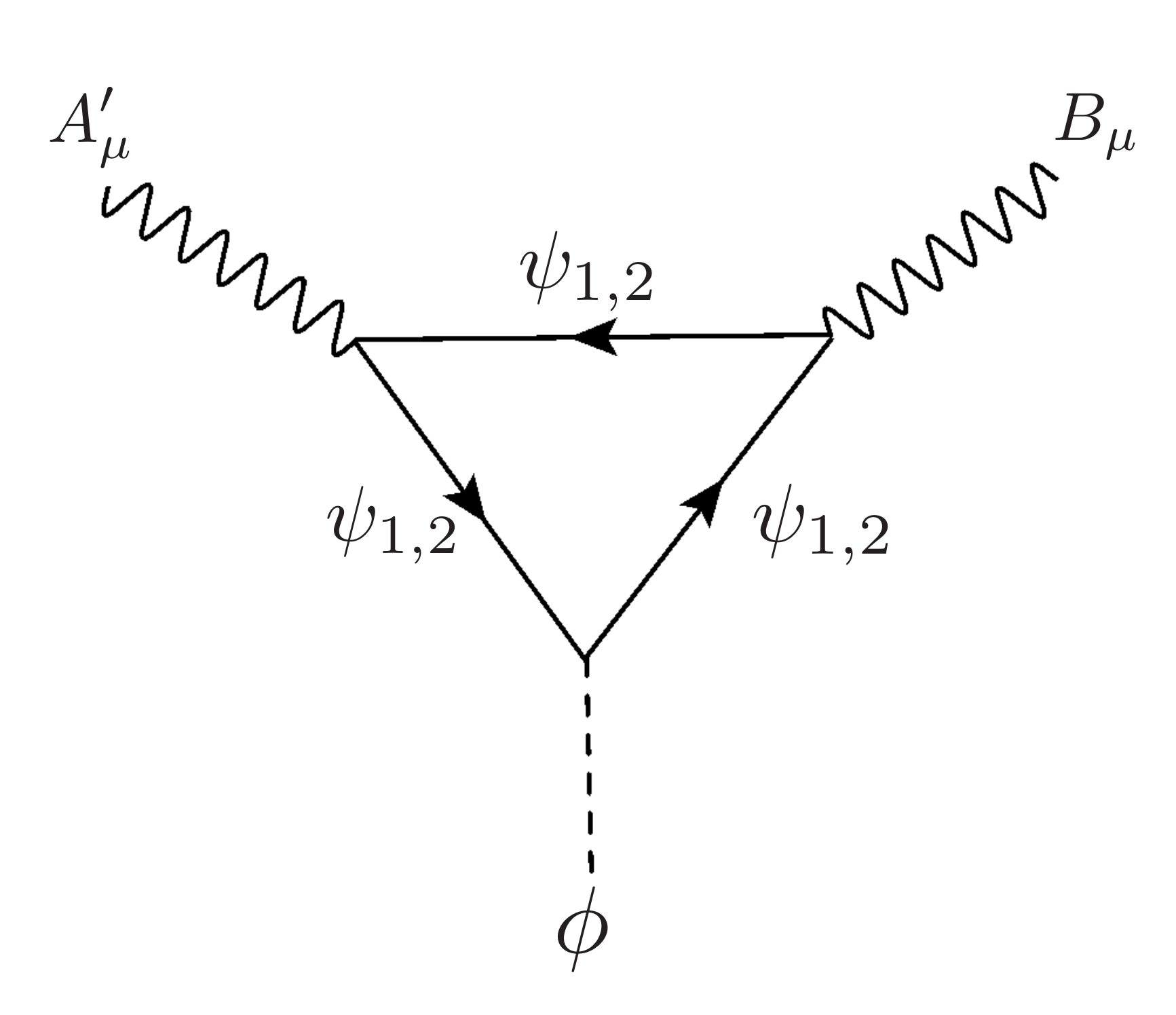}
 \end{center}
 \vspace{-10pt}
\caption{Diagrams contributing to Eq.~\eqref{eq:matching_Lambda'}.}
\label{fig:Lambda'diagram}
\end{figure}

The $\phi B^{\mu\nu} F'_{\mu\nu}$ operator is generated at 1-loop from $\psi_{1,2}$ loops (see Fig.~\ref{fig:Lambda'diagram}).
We find
\begin{align}
\frac{1}{\Lambda'} = \frac{y' g' Y_\psi g_Y^{\phantom*}}{3\pi^2 M'}
\,.\label{eq:matching_Lambda'}
\end{align}
Then, following the steps that led to the bounds~\eqref{eq:rough:M':Lambda-Lambda'} 
with more precise numerical coefficients for this model,
we find
\begin{align}
M' \lsim 2 \times 10^2\>\GeV \>  g' Y_\psi \sqrt{\frac{\Lambda' / \Lambda}{10^4}}
\!\left( \frac{\nc}{4 \times 10^3\>\keV^3} \right)^{\!1/4}
\,.\label{eq:Mpsi}
\end{align}
It should be emphasized that  this inequality is still rough as it comes from the condition not to fine-tune $\lambda_\phi$ against quantum corrections, which is not a quantitatively sharp criterion.
Moreover, the right-hand side above will be enhanced by a factor of $N^{3/4}$ if there are $N$ pairs of $\psi_{1,2}$, 
which is natural if $\psi_i$ forms a multiplet under some other symmetry such as SU(2)$_W$.

Let us now explicitly UV complete the electron density dependent operator.
The aforementioned requirement of generating the operator at tree level via $O(1)$ couplings to the electron and/or Higgs
strongly suggests vectorlike leptons with Yukawa couplings to $\phi$.
A minimal possibility is to consider two vector-like pairs of leptons, $L_{\rm L, R}$ and $E_{\rm L, R}$, with masses $M_L$ and $M_E$, respectively.
All of these are neutral under U(1)$_{A'}$, 
and they are all odd under $Z_2^{A'}$.
Writing down all gauge and $Z_2^{A'}$ invariant renormalizable interactions, we have
\begin{align}
{\cal L} 
\supset&\>  
-y_E^{\phantom*} H \bar L_{\rm L}  E_{\rm R} 
- y_E' H \bar L_{\rm R} E_{\rm L} 
- y_1 \, \phi \bar \ell_{\rm L} L_{\rm R} - y_2 \, \phi \bar e_{\rm R} E_{\rm L}
\nonumber\\
&\>+ \text{h.c.}
\,,\label{eq:vectorlike}
\end{align}
where $\ell_{\rm L}$ and $e_{\rm R}$ are the SM lepton fields.
Integrating out the vector-like leptons at tree level, we find
\bea
\frac{y_e}{\Lambda^2} 
= \frac{2y_E^{\phantom*} y_1 y_2^*}{M_L M_E}
\,.\label{eq:matching1}
\eea
Therefore, $c = 2y_E^{\phantom*}$, $y = \sqrt{y_1 y_2^*}$, and $M = \sqrt{M_L M_E}$ in Eq.~\eqref{eq:rough:matching:M}.
Upon integrating out $L$ and $E$, corrections to $\lambda_\phi$ is given by
\begin{align}
\Delta \lambda_\phi 
= -\frac{3}{2\pi^2} 
   \!\left( 2|y_1|^4 \ln\frac{M_L}{\mu}
           +|y_2|^4 \ln\frac{M_E}{\mu}
    \right),     
\end{align}
up to scheme dependent finite terms.
So, our estimate~\eqref{eq:rough:M:Lambda-Lambda'} is now refined as
\begin{align}
\sqrt{M_L M_E} \lsim 1\>\TeV \> \sqrt{y_E^{\phantom*}} \, \sqrt{\frac{10^4}{\Lambda' / \Lambda}}
\!\left( \frac{\nc}{4 \times 10^3\>\keV^3} \right)^{\!1/4}
\,.\label{eq:M:Lambda-Lambda'}
\end{align}
Thus, this can be experimentally probed as well. 
Unlike the guaranteed existence of new hypercharged particles at $M'$,
however, the vectorlike leptons may not be a unique possibility.
For example, it may be possible that the new particles at $M$ carry no SM charges,
e.g., via some kind of a Higgs portal.
Also, a viable range of $\Lambda' / \Lambda$ is small in the simple example construction above, 
which could become larger for a complicated model with multiple hierarchical scales and couplings.
But such model building exercises are beyond the scope of our present work.

The simple UV model above offers rich experimental probes. 
First, electroweak/Higgs precision tests place severe constraints on a vector-like lepton~\cite{Ellis:2014dza}, which tends to push the mass beyond the LHC reach~\cite{Kumar:2015tna, Aaboud:2019trc,Khachatryan:2016sfv}. 
However, if the flavor/CP structures of the Yukawa couplings~\eqref{eq:vectorlike} are generic, 
the vector-like-lepton sector can be indirectly tested through lepton flavor violation, such as $\mu \to e\gamma$ and $\mu\to eee$~\cite{Papa:2020umc}, as well as the electric dipole moment of the electron~\cite{Andreev:2018ayy}.

On the other hand, $\psi_i$ ($i=1,2$) can be produced at the LHC because the mass scale is expected to be a few hundred GeV as in~\eqref{eq:Mpsi}.
They can be constrained by searches for heavy charged states~\cite{Aaboud:2019trc,Khachatryan:2016sfv}. 
The bound can be relaxed if $\psi_i$ is a weak multiplet containing a neutral state $\psi_i^0$ (e.g., a weak doublet (triplet) with $Y_\psi = 1/2$ ($Y_\psi=1$)), 
which will allow weak decays of the charged states, e.g., $\psi_i^\pm \to \psi_i^0 \pi^\pm$.
The neutral states $\psi_i^0$ can be a WIMP dark matter candidate as they are stable thanks to the accidental U(1)$_{\psi_1}\times$U(1)$_{\psi_2}$ symmetry mentioned below~\eqref{eq:y'_Yukawa}.
In this case, searches for disappearing tracks will be relevant (e.g.~\cite{Sirunyan:2020pjd, ATL-PHYS-PUB-2019-011}). 
Detailed phenomenologies are UV model specific and beyond the scope of the present paper.

\section{Comparisons with other similar scenarios}
\label{sec:difficulties}
The key idea of our dark photon model is that the kinetic mixing parameter is density-dependent.
Similar ideas may be applied in different ways.
For example, one might consider a density dependent mass instead of coupling.
Or, as originally discussed in Refs.\,\cite{Bloch:2020uzh, DeRocco:2020xdt}, an axion-like particle (ALP) may be the signal particle corresponding to the XENON1T excess.
In the following, we would like to discuss various other scenarios, including both new and existing proposals, and point out difficulties that arise in each case to highlight the success of the dark photon model.

\subsection{Alternative dark photon scenarios}
The most obvious alternative to our dark photon scenario of Sec.~\ref{sec:dark} would be the following.
In Fig.~\ref{fig:bestfit}, instead of going down by turning off $\epsilon$ in dense stars, going far to the left by turning off $\mAd$ in dense stars could also evade the HB/RG bounds.
It is straightforward to modify the Lagrangian~\eqref{eq:Lagrangian} to realize this alternative. 
We just need to do two replacements: 
(i) replace the $\phi B^{\mu\nu} F'_{\mu\nu}$ interaction by a bare kinetic mixing term $\epsilon B^{\mu\nu} F'_{\mu\nu}$ with a {\it fixed} mixing parameter $\epsilon = 2.5 \times 10^{-13}$,
and (ii) replace the real scalar $\phi$ by a complex scalar $\Phi$ charged under U(1)$_{A'}$, and delete the St\"uckelberg mass term for $A'$. 
Then, denoting the U(1)$_{A'}$ gauge coupling by $g'$, 
we would have $\mAd = g'\langle\Phi\rangle = 2.3\>\keV$ in the vacuum/sun while $\mAd = 0$ in HB/RG stars.  
This would be a simpler theory as it introduces only one non-renormalizable operator, $m_e \bar{e} e |\Phi|^2 / \Lambda^2$, rather than two.

It does not work, however.
Essentially, the problem is that, unlike $\phi$ in the model of Sec.~\ref{sec:dark}, $\Phi$ here is charged under U(1)$_{A'}$.
The CW potential from $A'$ loop can then more directly impact the flatness of the $\Phi$ potential here than in the $\phi$ case.
If we decrease $g'$ to mitigate the impact, 
we would have to increase $\langle \Phi \rangle$ to keep $\mAd$ fixed in the vacuum/sun, which, however, would require an even more flat potential.
While the model still passes all 1-loop tests, it does not survive through the 2-loop diagrams discussed around Eq.~\eqref{eq:e-h-loop}.
Here, Eq.~\eqref{eq:e-h-loop} should be adapted as follows.
The scale $\Lambda$ can be eliminated in favor of $v_0$ and $\lambda_\phi$ via $\lambda_\phi v_0^2 \sim m_e \nc / \Lambda^2$, where $v_0$ can be further expressed as $v_0 = \mAd / g'$.
The modified Eq.~\eqref{eq:e-h-loop} now roughly reads $g'^4 / \lambda_\phi \gsim (m_h \mAd^2 / \nc)^2 / (16\pi^4) \sim 10^7$ for $\mAd \sim 2\>\keV$ and $\nc \sim 10^3\>\keV^3$.
But this contradicts with the condition $g'^4 / \lambda_\phi \lsim 1$ to ensure the 1-loop CW potential from $A'$ loop not to destroy the flat potential.

The same reason also excludes yet another dark photon scenario in which $\mAd$ {\it increases} in dense stars instead of decreasing. 
This is the same model as above except that the sign of the $\bar{e} e |\Phi|^2 / \Lambda$ term is flipped so that U(1)$_{A'}$ is more broken in denser stars than in the vacuum/sun.
Since the estimates above never relied on the sign of this term, this alternative is also unattainable due to too large quantum corrections. 
We thus conclude that, at least among the simplest dark photon models, our scenario of Sec.~\ref{sec:dark} seems the only viable option.

\subsection{Density depedent ALP models}
We now consider the case when an ALP gives the XENON1T signal instead of a dark photon.
In this case, we have two directions for utilizing the density dependence: suppressing ALP couplings or increasing the ALP mass in dense stars.

Let us begin by the idea of suppressing ALP couplings.
In the ``chameleon-like'' ALP model of Ref.~\cite{Bloch:2020uzh}, there is an operator, $c_{ee} m_e \bar{e} e XS / M^2$, where $X$ and $S$ are respectively real and complex scalars, both of them being odd under a $Z_2$ symmetry.
The ALP, $a$, comes from $S \sim (f_a + s) \, {\rm e}^{{\rm i} a / f_a}$.
The potential $V(X)$ is identical to our potential $V(\phi)$ in~\eqref{eq:Vphi} with the replacements of $\phi$ and $m_e \bar{e} e / \Lambda^2$ by $X$ and $\rho / M^2$, respectively, where $\rho$ is the {\it energy} density as opposed to the number density.
Thus, in dense stars $\langle X \rangle$ vanishes and eliminates the stellar cooling problem due to ALP production.

Unfortunately, as it is already discussed in Ref.~\cite{Bloch:2020uzh}, the flatness of the potential suffers from too large quantum corrections. 
In addition, since $\rho$ appears in a quantum field theory Lagrangian, we interpret it as arising as the expectation value of a dynamical operator such as $m_e \bar{e} e$, $m_N \bar{N} N$ or ${\rm tr}[G_{\mu\nu} G^{\mu\nu}]$, where $N$ and $G$ are a nucleon and the gluon field, respectively.
Then, there are additional loops from, e.g., a gluon loop or two-loop diagrams as in Fig.~\ref{fig:e-h-diagram} left. 
We also find that the  $c_{ee}$ operator  generates a tadpole term of $X$ in dense stars proportional to $n_e$ which prevents $\langle X \rangle$ from vanishing.
These effects could lead to even stronger constraints on the potential. 

We now switch to the idea of increasing the ALP mass. 
In Ref.\,\cite{DeRocco:2020xdt}, the ALP mass increases in dense stars as $\langle \phi \rangle$ 
and its production is kinematically forbidden 
when the mass becomes heavier than the core temperature of the star.
The ALP is the ``pion'' of a dark QCD, where the dark quark masses are proportional to $\langle \phi \rangle^2$ via a see-saw with a heavy fermion.
A nonzero $\langle \phi \rangle$ arises from a tadpole due to a Yukawa interaction $\phi \bar{N} N$ when the nucleon density $\bar{N} N$ is turned on.

Ref.\,\cite{DeRocco:2020xdt} already mentions fine tuning issues in some parameters in the dark sector.
Here, to relate to our earlier discussions, let us look at 1-loop quantum corrections to the $\phi$ potential.
In particular, the see-saw requires Yukawa couplings of $\phi$ to a light dark quark and a heavy dark quark, 
which then contribute to the 1-loop CW potential from dark quark loops,
where the coefficient of $\phi^{2n}$ is roughly given by $y^{2n} M^{4-2n}/16\pi^2$.
Requiring the generated $\phi^{2n}$ term to be smaller than the tree-level $\lambda \phi^4$ for $\phi \sim \langle \phi \rangle$, 
we have $y^{2n} (\langle \phi \rangle / M)^{2n-4}/16\pi^2 \lsim \lambda$. 
In the benchmark model of Ref.\,\cite{DeRocco:2020xdt}, 
we have $\lambda = 10^{-78}$, $y = \sqrt{g_1 g_2} = 3 \times 10^{-11}$, $M = 30\,\TeV$,
and $\langle \phi \rangle \simeq 3 \times 10^{12} \, m_a$ 
with $m_a \simeq 7\>\MeV$ in white dwarfs.
With these values, the potential needs to be fine-tuned up to and including $\phi^{10}$.

As our final example, we consider a scenario where the ALP coupling to the electron is proportional to $1 / \langle \phi \rangle$, 
where $\langle \phi \rangle$ increase in dense stars to suppress the ALP production.
This can be realized in a DFSZ-like model of the axion~\cite{Dine:1981rt, Zhitnitsky:1980tq} where $\langle\phi\rangle$ breaks the Pecci-Quinn (PQ) symmetry 
in a density dependent way as in~\eqref{eq:Vphi} but with the sign of the $\Lambda^2$ term being flipped.
Provided that $\langle \phi \rangle$ in the vacuum, $v_0$, is much larger than the electroweak symmetry breaking scale, 
the coupling of the axion to the electron is given by
$(m_e / v_0) \, a \bar{e} \gamma_5 e$.
Then, from the value of $g_{aee}$ in Ref.~\cite{Aprile:2020tmw}, we get $v_0\sim 10^9\,\GeV$.

Now, the cutoff $\Lambda$ of the effective theory provides the maximum possible value of $\langle \phi \rangle$ in a dense environment.
The validity of the theory in white dwarfs, where the electron density is $10^4$ higher than in the sun, requires $\Lambda \gsim 10^2\,v_0 \sim 10^{11}\,\GeV$.
Such large $\Lambda$ forces $\mu_\phi \lsim 10^{-11}\,{\rm eV}$. 
The large hierarchy between $m_\phi$ and $v_0$ requires a tiny quartic coupling $\lambda_\phi \sim \mu_\phi^2 / v_0^2 \lsim 10^{-58}$.
But the theory needs an interaction of the form $c \phi^2 H_1 H_2$ to identify the global U(1) carried by $\phi$ as the U(1)$_\text{PQ}$ carried by the two Higgs doublets (and, therefore, the SM fermions).
This operator radiatively corrects $\lambda_\phi$, which is roughly  $c^2/16\pi^2$. But one of the CP-odd scalars, $A$, has a mass $m_A \sim  \sqrt{c} v_0$. Requiring $m_A \gsim 100\>\GeV$, we have $c \gsim 10^{-14}$, but this implies the above correction to $\lambda_\phi$ is at least $\sim 10^{-30}$, which is $\gg 10^{-58}$. 
Thus, again, the model suffers from too large quantum corrections.

To conclude, it appears to be generic that an extremely flat potential is needed to explain the XENON1T excess while evading the stellar cooling problems in denser stars by a density-dependent vev of a scalar field.
What we have seen in the examples above is that it is not easy to maintain the flatness of the potential at quantum level.
Therefore, it is quite nontrivial that our dark photon model has a large region in the parameter space where the potential receives no large quantum corrections (except for the mass terms, of course).

\section{Summary}
We have shown that a solar dark photon with a density-dependent kinetic mixing with the photon could explain the XENON1T signal excess while avoiding bounds from stellar cooling.
We have provided an explicit effective Lagrangian in which the kinetic mixing is proportional to the expectation value of a scalar field $\phi$ with an electron density dependent potential.
The kinetic mixing vanishes in dense stars, permitting the model to evade stellar cooling bounds, while acquires a nonzero value in the sun/vacuum, providing the signal strength and shape to explain the XENON1T excess.

We have studied both phenomenological and theoretical constraints.
We have found that stellar cooling due to $\phi$-pair and $\phi$-$A'$ productions provides the lower bounds on the scales $\Lambda$ and $\Lambda'$ of the effective theory.
As these bounds imply an extremely flat potential of $\phi$, 
we have studied the impact of radiative corrections on the flatness.
This leads to strong lower and upper bounds on the ratio of $\Lambda'$ to $\Lambda$.
All these phenomenological and theoretical constraints are summarized in Fig.\,\ref{fig:moneyplot}.
We emphasize that having a large allowed parameter space is quite nontrivial without fine tuning against quantum corrections.

We have presented a general argument that the scales in our effective theory predicts new electrically charged particles with roughly sub-TeV masses in the UV completion.
Our scenario is thus testable via laboratory experiments such as lepton flavor/CP violation measurements, electroweak/Higgs precision tests, and colliders.
We have constructed a simple explicit example of the UV completion in terms of fermions charged under $U(1)_Y\times U(1)_{A'}$ and vectorlike leptons.
This simple model can contain a dark matter candidate and provide a rich collider phenomenology with displaced vertices and disappearing tracks.

{\it Note Added---}
In the final stage of writing this paper, new work on the solar dark photon \cite{Lasenby:2020goo} appeared. If there is an additional flux of dark photons as pointed out in Ref~\cite{Lasenby:2020goo}, the best fit point of the kinetic mixing for the XENON1T excess will decrease from the current benchmark $\epsilon\simeq 2.5\times10^{-13}$ to $\epsilon\sim 10^{-14}$. This will still be in tension with the cooling bounds by HB/RG stars. Therefore, the main points of our paper will be unchanged.

\section*{Acknowledgment}
This work is supported by the US Department of Energy grant DE-SC0010102. 
KT thanks to Natsumi Nagata, Kazunori Nakayama, Kenichi Saikawa, and Diego Redigolo for useful discussions.  

\bibliographystyle{utphys}
\bibliography{XENON1Tbib}

\end{document}